\documentclass[prl,twocolumn,showpacs,showkeys,amsmath]{revtex4}
\usepackage{bm,graphicx}
\renewcommand{\d}{\mathrm{d}}
\newcommand{\e}{\mathrm{e}}
\renewcommand{\i}{\mathrm{i}}
\begin{document}

\title{Huygens-Fresnel principle for $N$-photon states of light}

\date{\today}

\author{E. Brainis}
\email{ebrainis@ulb.ac.be}
\affiliation{Service OPERA, Universit\'e libre de Bruxelles, Avenue F. D. Roosevelt 50, B-1050 Bruxelles, Belgium}

\begin{abstract}
We show that the propagation of a $N$-photon field in space and time can be described by a generalized Huygens-Fresnel integral. Using two examples, we then demonstrate how familiar Fourier optics techniques applied to a $N$-photon wave function can be used to engineer the propagation of entanglement and to design the way the detection of one photon shapes the state of the others.  
\end{abstract}


\maketitle

In quantum optics and quantum information science, one often deals with systems in which the total photon number $N$ is known, the challenge being to prepare the required $N$-photon state and engineer its evolution. When $N$ is small, a wave-function formalism provides a more compact and intuitive physical description than the usual quantum field formalism. This has been advocated in the recent years \cite{Smith2006,Smith2007} and justifies the renewed interest in photon wave-mechanic. 

Single photons propagate in space and diffract on obstacles exactly as classical waves do. How a $N$-photon wave function propagates is less obvious. In Ref.~\onlinecite{Smith2006}, the authors used a two-photon Maxwell-Dirac equation to study the disentanglement of a photon pair. In this Letter, we would like to lay the grounds of a propagation theory of $N$-photon wave-packets. Instead of using differential equations, we generalize the Huygens-Fresnel principle to obtain an \emph{integral} formulation relating the values of the $N$-photon wave function at a given time to its values on a fixed reference surface (usually a plane) at previous times. Familiar Fourier optics techniques can then be applied in order to engineer the propagation of entanglement and shape the photon state through the detection process. We illustrate this point using two examples. Our approach generalizes previous works on the ``ghost imaging'' properties of photon pairs produced by parametric down conversion \cite{Abouraddy2002,Shimizu2006} and is relevant to many applications of modern quantum optics, including quantum super-resolution imaging, quantum lithography, as well as spatial quantum communication.

The very idea of using \emph{position} wave functions to describe the state of $N$-photon systems relies on the existence (but not uniqueness) of a photon position operator $\hat{\mathbf{r}}$, the Cartesian components of which are commuting Hermitian operators satisfying $[\hat{r}_k,\hat{p}_l]=i\hbar \delta_{kl}$, $\hat{\textbf{p}}$ being the photon momentum \cite{Hawton1999a,Hawton2001}. The eigenfunctions of $\hat{\mathbf{r}}$ are transverse waves that can be interpreted as localized-photon states \cite{Hawton1999}. Any admissible single-photon wave function is obtained as a linear combination of these localized states. $N$-photon wave functions are symmetric elements of the tensor product of $N$ single-particle Hilbert spaces. Because the definition of the position operator $\hat{\mathbf{r}}$ is not unique, there is more than one way to assign a position wave function to a single photon. The most popular one is probably the so-called Bialynicki-Birula-Sipe wave function $\bar{\psi}(\mathbf{r},t)=[\bm{\psi}_+(\mathbf{r},t) \ \bm{\psi}_-(\mathbf{r},t)]$ which has two vector components corresponding to photons with positive and negative helicity \cite{Bialynicki-Birula1994,Bialynicki-Birula1996a}. Each vector component has a Fourier expansion that reads
\begin{equation}\label{psih}
\bm{\psi}_\pm(\mathbf{r},t)=\int \d^3 k \ \sqrt{\hbar k c} \ \mathbf{e}_\pm(\mathbf{k}) \ f_\pm(\mathbf{k}) \ \frac{\e^{\i (\mathbf{k}\cdot \mathbf{r}- kc \ t)}}{(2\pi)^{3/2}},
\end{equation}
where $k=|\mathbf{k}|$ and $\mathbf{e}_\pm(\mathbf{k})$ are the unit circular polarization vectors for photons propagating in the $\mathbf{k}$-direction. Normalization is such that the complex coefficients $f_\pm$ satisfy $\sum_{h=\pm}\int \d^3 k \ |f_h(\mathbf{k})|^2=1$. This wave function transforms as an elementary object under Lorentz transformation and can be easily connected to Maxwell fields. In this Letter, we write the wave function is a slightly different (but equivalent) way that consists in summing both helicity components together:
\begin{equation}\label{Psi}
\bm{\Psi}(\mathbf{r},t)=\bm{\psi}_+(\mathbf{r},t) + \bm{\psi}_-(\mathbf{r},t).
\end{equation}
This provides a vector representation instead of the bi-vector one \cite{Sipe1995}. Since $\bm{\psi}_+$ and $\bm{\psi}_-$ are orthogonally polarized, they never mix: if $\bm{\Psi}$ is given,  
$\bm{\psi}_+$ and $\bm{\psi}_-$ can be deduced. Therefore the information content in the vector function $\bm{\Psi}$ is the same as in the bi-vector field $\bar{\psi}$.

Working with wave functions (`first quantization'' formalism) is fully equivalent to working with quantum fields (``second quantization'' formalism). Replacing the complex coefficients $f_{\pm}(\mathbf{k})$ by annihilation operators $\hat{a}_{\pm}(\mathbf{k})$ in Eq.~(\ref{psih}), the fundamental quantum field is found to be proportional to the \emph{positive frequency} part of the electric field:
\begin{eqnarray*}
\hat{\bm{\Psi}}(\mathbf{r},t)&=& \sum_{h=\pm}\int \d^3 k \ \sqrt{\hbar k c} \ \mathbf{e}_h(\mathbf{k}) \ \hat{a}_h(\mathbf{k}) \ \frac{\e^{\i (\mathbf{k}\cdot \mathbf{r}- kc \ t)}}{(2\pi)^{3/2}} \nonumber \\
&= & -\i \ \sqrt{2\epsilon_0} \ \hat{\mathbf{E}}^{(+)}(\mathbf{r},t).
\end{eqnarray*}  
Note that $\hat{\bm{\Psi}}\propto \hat{\mathbf{E}}^{(+)}$ holds information about \textit{both} electric and magnetic field. Therefore, it provides a complete information about electromagnetic configuration. To show this explicitly, one decomposes  $\hat{\bm{\Psi}}$ again into its helicity components $\hat{\bm{\psi}}_+$ and $\hat{\bm{\psi}}_-$ and subtract them: this yields  $\hat{\mathbf{B}}^{(+)}=\sqrt{\mu_0/2}(\hat{\bm{\psi}}_+-\hat{\bm{\psi}}_-)$, the positive frequency part of the magnetic field.
In the second quantization formalism, the state of a single-photon wave packet writes: $|\Psi \rangle=\sum_{h=\pm}\ \int \d^3k \ f_h(\mathbf{k}) \ |1_{\mathbf{k},h}\rangle$, where $|1_{\mathbf{k},h}\rangle=a_h^{\dag}(\mathbf{k})|0\rangle$ and $f_\pm(\mathbf{k})$ are the same spectral amplitudes that appear in Eq.~(\ref{psih}). The connection between the first and second quantization formalism is given by the relation
$\bm{\Psi}(\mathbf{r},t)=\langle 0 |\hat{\bm{\Psi}}(\mathbf{r},t)|\Psi \rangle=-\i \ \sqrt{2\epsilon_0} \ \langle 0 |\hat{\mathbf{E}}^{(+)}(\mathbf{r},t)|\Psi \rangle$.
Since $\langle \Phi |\hat{\bm{\Psi}}(\mathbf{r},t)|\Psi \rangle=0$ for all $|\Phi\rangle\neq |0\rangle$, we have $\Psi^*_{i'}(q')\Psi_i(q) = \langle \Psi |\hat{\Psi}_{i'}^\dag(q') \hat{\Psi}_i(q)|\Psi \rangle= 2\epsilon_0 \langle \hat{E}_{i'}^{(-)}(q')  \hat{E}_i^{(+)}(q)\rangle$ for any pair of points $q=(\mathbf{r},t)$ and $q'=(\mathbf{r}',t')$,
where the indexes  $(i,i')\in \{x,y,z\}^2$ represent Cartesian components. This relates the Bialynicki-Birula-Sipe wave-function to the usual first-order correlation functions of coherence theory. In particular, $|\bm{\Psi}(\mathbf{r},t)|^2$ is proportional to the probability to detect the photon energy at point $\mathbf{r}$ at time $t$. This gives to the Bialynicki-Birula-Sipe wave function the usual interpretation of a probability amplitude to find the photon energy at some position. 

The generalization to $N$-photon states 
\begin{multline*}
|\Psi\rangle = \sum_{h_1,\ldots,h_N} \int \d^3k_1 \ldots \int\d^3k_N \
  \\ f_{h_1,\ldots,h_N}(\mathbf{k}_1,\ldots,\mathbf{k}_N) |1_{\mathbf{k}_1,h_1},\ldots,1_{\mathbf{k}_N,h_N} \rangle
\end{multline*}
is straightforward. The connection between wave functions and fields is given by
\begin{multline}\label{amplN}
\Psi_{i_1\ldots i_N}(q_1,\ldots, q_N)=(-\i)^N (2\epsilon_0)^{N/2}\\
 \times \langle 0 |\hat{E}^{(+)}_{i_N}(q_N)\ldots \hat{E}^{(+)}_{i_1}(q_1)|\Psi \rangle
\end{multline}
and
\begin{multline}\label{cohN}
\Psi^*_{i'_1\ldots i'_N}(q'_1,\ldots, q'_N) \Psi_{i_1\ldots i_N}(q_1,\ldots, q_N)=(2\epsilon_0)^N \\
\times \langle \hat{E}_{i'_1}^{(-)}(q'_1)\ldots\hat{E}_{i'_N}^{(-)}(q'_N) \hat{E}_{i_N}^{(+)}(q_N) \ldots \hat{E}_{i_1}^{(+)}(q_1)\rangle.
\end{multline} 
Eq.~(\ref{cohN}) shows that \emph{any} field correlation function of a $N$-photon system can be computed as a product of two tensor elements of the $N$-photon wave function. The best way to compute the propagation of the $N$-photon wave function may depend on the situation. Often, the photonic state is prepared in such a way that the wave function is known at \emph{all} times, but only on a specific surface $\Sigma$. Therefore, a diffraction theory of $N$-photon states is needed. An important example is the generation of entangled photon pairs in a nonlinear crystal. In that case, $\Sigma$ is the output face of the crystal and diffraction from that plane leads to phenomena the phenomena of ``ghost imaging'' \cite{Strekalov1995}.

\begin{figure}[b]
\centerline{\includegraphics[width=8.6cm]{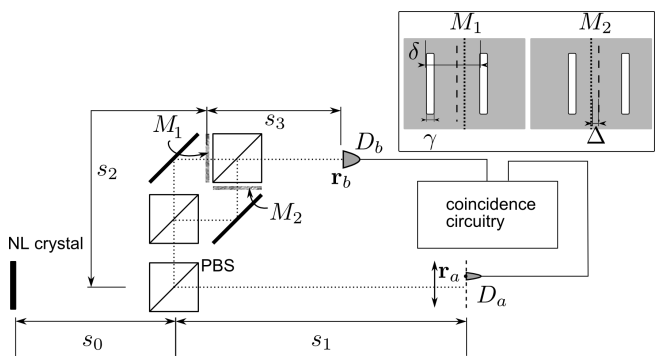}}
\caption{Interference effect in quantum imaging with two separated diffraction masks. \label{fig:1}}
\end{figure}

To understand how $N$-photon detection correlations spread in space and time, we first consider free-space propagation. We also make the simplifying assumption that we deal with paraxial states of light, in which case polarization does not change much during propagation. Therefore, we drop the polarization-related indexes. Considering photons propagating along the $z$-axis, we use the Huygens-Fresnel principle \cite{Goodman2005} to express $\hat{E}^{(+)}$ at some point as a function of its values on the reference plane $\Sigma$ at $z$-coordinate $\zeta$,
\begin{equation}\label{HF}
\hat{E}^{(+)}(\mathbf{r},t)=\frac{1}{2\pi c }\iint \d^2\rho^\perp \frac{\frac{\d}{\d t}\hat{E}^{(+)}(\bm{\rho},t-\frac{|\mathbf{r}-\bm{\rho}|}{c})}{|\mathbf{r}-\bm{\rho}|} ,
\end{equation}
and inject this in Eq.~(\ref{amplN}). This yields
\begin{multline}\label{GHF}
\Psi(\mathbf{r}_1,t_1,\ldots,\mathbf{r}_N,t_N)=\frac{1}{(2\pi c)^N}\iint \d^2\rho_{1}^\perp\ldots\iint \d^2\rho_{N}^\perp\\
 \frac{\frac{\d}{\d t_1}\cdots\frac{\d}{\d t_N}\Psi(\bm{\rho}_1,t_1-\frac{|\mathbf{r}_1-\bm{\rho}_1|}{c},\ldots,\bm{\rho}_N,t_N-\frac{|\mathbf{r}_N-\bm{\rho}_N|}{c})}{|\mathbf{r}_1-\bm{\rho}_1|\cdots |\mathbf{r}_N-\bm{\rho}_N| }.
\end{multline}
We call this integral the generalized Huygens-Fresnel (GHF) principle for $N$-photon wave functions. In Eqs.~(\ref{HF}) and (\ref{GHF}), $\bm{\rho}_j=(\xi_j,\eta_j,\zeta)$ ($j\in\{1,\ldots,N\}$) are points in the $\zeta$-plane and $\bm{\rho}_j^\perp=(\xi_j,\eta_j)$ are their transverse components. In the optical domain, photons can usually be  considered as quasi-monochromatic. Therefore, the wave function can be written 
$\Psi(\mathbf{r}_1,t_1,\ldots,\mathbf{r}_N,t_N)=a(\mathbf{r}_1,t_1,\ldots,\mathbf{r}_N,t_N) \exp{(-\i 2\pi c (\frac{ t_1}{\lambda_1}+\cdots +\frac{ t_N}{\lambda_N}))}$,
where $a(\mathbf{r}_1,t_1,\ldots,\mathbf{r}_N,t_N)$ is a slowly varying function of time and $\lambda_j$ ($j\in\{1,\ldots,N\}$) are the central wavelengths of the photons. Note that nothing prevents  photons from having the same central wavelength or even being indistinguishable. Inserting that anzats in Eq.~(\ref{GHF}) and taking into account that $a(\mathbf{r}_1,t_1,\ldots,\mathbf{r}_N,t_N)$ is slowly varying in time, one obtains
\begin{multline}\label{GHF2}
a(\mathbf{r}_1,t_1,\ldots,\mathbf{r}_N,t_N)=\frac{(-\i)^N}{\lambda_1\ldots\lambda_N}\iint \d^2\rho_{1}^\perp\ldots\iint \d^2\rho_{N}^\perp\\
a(\bm{\rho}_1,t_1-\frac{|\mathbf{r}_1-\bm{\rho}_1|}{c},\ldots,\bm{\rho}_N,t_N-\frac{|\mathbf{r}_N-\bm{\rho}_N|}{c})\\
\frac{\exp\left({\i \frac{2 \pi}{\lambda_1} |\mathbf{r}_1-\bm{\rho}_1|}\right)}{|\mathbf{r}_1-\bm{\rho}_1|}\ldots \frac{\exp\left({\i \frac{2\pi}{\lambda_N} |\mathbf{r}_N-\bm{\rho}_N|}\right)}{|\mathbf{r}_N-\bm{\rho}_N|}.
\end{multline}
Eq.~(\ref{GHF2}) is only valid in free space. If propagation from $\bm{\rho}_i$ to $\mathbf{r}_i$ is through an optical system, the free space propagator
\begin{equation}
h_{fs}(\mathbf{r}_i,\bm{\rho}_i)=\frac{-\i}{\lambda_i} \ \frac{\exp\left({\i \frac{2 \pi}{\lambda_i} |\mathbf{r}_i-\bm{\rho}_i|}\right)}{|\mathbf{r}_i-\bm{\rho}_i|}
\end{equation}
must be replaced by the appropriate one $h_i(\mathbf{r}_i,\bm{\rho}_i)$. With this generalization, Eq.~(\ref{GHF2}) becomes
\begin{multline}\label{GHF3}
a(\mathbf{r}_1,t_1,\ldots,\mathbf{r}_N,t_N)=\iint \d^2\rho_{1}^\perp\ldots\iint \d^2\rho_{N}^\perp\\
a(\bm{\rho}_1,t_1-\frac{l(\mathbf{r}_1,\bm{\rho}_1)}{c},\ldots,\bm{\rho}_N,t_N-\frac{l(\mathbf{r}_N,\bm{\rho}_N)}{c})\\
h_1(\mathbf{r}_1,\bm{\rho}_1) \ldots h_N(\mathbf{r}_N,\bm{\rho}_N),
\end{multline}
where $l(\mathbf{r}_i,\bm{\rho}_i)$ is the optical path length from $\bm{\rho}_i$ to $\mathbf{r}_i$. Formula (\ref{GHF3}) assumes that there is only \emph{one} optical path from $\bm{\rho}_i$ to $\mathbf{r}_i$. However, interferometers with arms having different path lengths can be placed between $\bm{\rho}_i$ and $\mathbf{r}_i$. To take this into account, we generalize (\ref{GHF3}) in the following way:
\begin{multline}\label{GHF4}
a(\mathbf{r}_1,t_1,\ldots,\mathbf{r}_N,t_N)=\iint \d^2\rho_{1}^\perp\ldots\iint \d^2\rho_{N}^\perp\\
\sum_{k_1,\ldots,k_N }a(\bm{\rho}_1,t_1-\frac{l_{k_1}(\mathbf{r}_1,\bm{\rho}_1)}{c},\ldots,\bm{\rho}_N,t_N-\frac{l_{k_N}(\mathbf{r}_N,\bm{\rho}_N)}{c})\\
h_1^{(k_1)}(\mathbf{r}_1,\bm{\rho}_1) \ldots h_N^{(k_N)}(\mathbf{r}_N,\bm{\rho}_N).
\end{multline}
The indexes $k_i$ label the different paths from $\bm{\rho}_i$ to $\mathbf{r}_i$.
We illustrate formula (\ref{GHF4}) using two examples.
Consider the system in Fig.~\ref{fig:1}.
The scheme is similar to the original ``ghost imaging'' scheme \cite{Strekalov1995}, with the difference that \emph{two} diffraction masks (called $M_1$ and $M_2$) are placed in the arms of a balanced Mach-Zenhder interferometer. Co-propagating energy-degenerated ($\lambda_1=\lambda_2\equiv\lambda$) entangled photon pairs are generated by parametric down-conversion in a nonlinear crystal and split using polarizing beam splitter (PBS). The generation process entangles the photons in position and time. The detector $D_b$ clicks whenever the photon going upward passes through $M_1$ or $M_2$ and is detected on axis. The diffraction pattern of the second photon is detected by moving the point-like detector $D_a$. Assuming a Gaussian pump and broadband phase-matching, the wave function amplitude at the output of the nonlinear crystal is well approximated by
\begin{multline*}
a(\bm{\rho}_1,t_1,\bm{\rho}_2,t_2)= \frac{\exp\left(-\frac{t_2^2}{4T^2}\right)}{(2\pi T^2)^{1/4}} \ \frac{\exp\left(-\frac{\xi_2^2+\eta_2^2}{4S^2}\right)}{(2\pi S^2)^{1/2}} \\
\times \delta(\bm{\rho}^\perp_1-\bm{\rho}^\perp_2) \ \delta(t_1-t_2),
\end{multline*}
where $T$ and $S$ are the time and space rms-half-width of the pump pulse. The propagation in the optical system can be computed using formula (\ref{GHF4}), with
\begin{eqnarray*}
h_{1}(\mathbf{r}_a,\bm{\rho}_1)&=&\frac{-\i}{\lambda} \ \frac{\exp\left[{\i \frac{2 \pi}{\lambda} |\mathbf{r}_a-\bm{\rho}_1|}\right]}{|\mathbf{r}_a-\bm{\rho}_1|},\\
h_{2}^{(k)}(\mathbf{r}_b,\bm{\rho}_2)&=&\frac{(-1)^{k}}{\sqrt{2}\lambda^2} \ \iint M_k(\mathbf{r}^{\perp}) \frac{\exp\left[{\i \frac{2 \pi}{\lambda} |\mathbf{r}-\bm{\rho}_2|}\right]}{|\mathbf{r}-\bm{\rho}_2|} \nonumber\\
&\times &\frac{\exp\left[{\i \frac{2 \pi}{\lambda} |\mathbf{r}_b-\mathbf{r}|}\right]}{|\mathbf{r}_b-\mathbf{r}|}  \ \d^2 r^{\perp}
\end{eqnarray*}
where $M_k(\mathbf{r}^{\perp})$ ($k\in\{1,2\}$) are the mask transfer functions. The GHF integral (\ref{GHF4}) simplifies if $4S^2 \gg \lambda (s_0+ \min(s_1,s_2))$ (i.e. the  wavefront curvature of the generated photons can be neglected) and if the far-field conditions $d^2/\lambda\ll s_3$ and $d^2/\lambda\ll(s_0+s_2)$ apply, where $d$ is the relevant length of the mask profile functions. Dropping constant and quadratic phase factors, one finds 
\begin{multline*}
a(\mathbf{r}_a,t_a,\mathbf{r}_b,t_b)= \exp\left[{\frac{t_b-(s_0+s_2)/c)}{4T^2}}\right] \delta(t_a-t_b-\tau) \\
\exp\left[\i Q\right] \sum_{k=1}^2 \tilde{M}_k\left[\frac{2\pi}{\lambda}\left(\frac{x_a}{L}+\frac{x_b}{s_3}\right),\frac{2\pi}{\lambda}\left(\frac{y_a}{L}+\frac{y_b}{s_3}\right)\right],
\end{multline*}
where $L=(2s_0+s_1+s_2)$, $\tau=(s_1-s_2-s_3)/c$, $\mathbf{r}_i=(x_i,y_i)$ and $\tilde{M}_k(\mathbf{k}^\perp)=\int M_k(\mathbf{r}^{\perp}) \ \exp[-\i \mathbf{k}^{\perp}\cdot\mathbf{r}^\perp] \ \d r^\perp $ is the Fourier transform of the mask function $M_k(\mathbf{r}^\perp)$. The exponential factor indicates that the detection of a photon in $D_b$ can only happen a propagation time $(s_0+s_2)/c$ after the photon has been created and within the initial time uncertainty $T$. The $\delta$-factor shows that the difference $\tau$ in the detection times of $D_a$ and $D_b$ is just due to different propagation distances from the crystal to the detectors, as expected for time-entangled photons. The wave-front curvature of both photons are related through the quadratic phase $Q=(\pi/\lambda) [(x_a^2+y_a^2)/(2s_0+s_1+s_2)+(x_b^2+y_b^2)/s_3]$. If detector $D_b$, placed on axis ($\mathbf{r}_b^\perp=0$), detects a photon a time $t_b=t_*$, the wave function amplitude of the photon travelling towards $D_a$ is automatically projected on  
\begin{equation*}
a(\mathbf{r}_a,t_a)= \delta\left(t_a-\left[t_*+\tau\right]\right)\exp(\i Q_a)\sum_{k=1}^2 \tilde{M}_k\left(2\pi\frac{x_a}{\lambda L},2\pi\frac{y_a}{\lambda L}\right),
\end{equation*}
with $Q_a=(\pi/\lambda)(x_a^2+y_a^2)/(2s_0+s_1+s_2)$. The system of Fig. \ref{fig:1} can be used to produce heralded single photons with on-demand or adaptive spatial profile. One can mimic a complex modulation (amplitude and phase) using phase modulators at $M_1$ and $M_2$. Such a system would be useful in the context of earth-satellite quantum cryptography \cite{Villoresi2008} if true single photons had to be used. Even with simple masks, such as the double slits in the inset of Fig. \ref{fig:1}, non trivial shaping can be done. If $2\Delta=\delta$, changing the interferometric phase $\phi$ from 0 to $\pi$ doubles the spatial frequency of the interference fringes seen by $D_a$. Single photons with very different spatial profiles can be created by tuning only one parameter. 
\begin{figure}[b]
\centerline{\includegraphics[width=8.6cm]{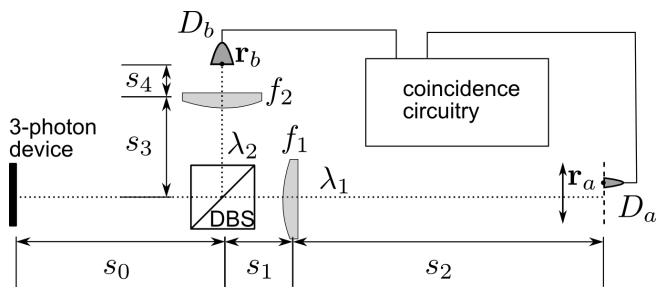}}
\caption{Generation of heralded linear superposition of localized two-photon states of light. \label{fig:2}}
\end{figure}
As a second example, consider the system in Fig. \ref{fig:2}. A nonlinear device produces photon triplets with time and space entanglement. (Such a device do not exist yet, but time-entangled triplets have been demonstrated recently \cite{Hubel2010}. A bulk version of that experiment would exhibit space entanglement as well.). We assume that, in the output plane of the device, the three-photon amplitude is
\begin{equation*}
\begin{split}
&a(\bm{\rho}_1,t_1,\bm{\rho}_2,t_2,\bm{\rho}_3,t_3)= \frac{\exp\left(-\frac{t_3^2}{4T^2}\right)}{(2\pi T^2)^{1/4}} \ \frac{\exp\left(-\frac{\xi_3^2+\eta_3^2}{4S^2}\right)}{(2\pi S^2)^{1/2}} \\
&\phantom{=}\times \delta(\bm{\rho}^\perp_1-\bm{\rho}^\perp_3) \ \delta(\bm{\rho}^\perp_2-\bm{\rho}^\perp_3) \ \delta(t_1-t_3) \ \delta(t_2-t_3).
\end{split}
\end{equation*}
We also assume that two photons have a common wavelength $\lambda_1$, while the third one has a different one $\lambda_2$. They are separated using the dichroic beam splitter (DBS). In the $\lambda_1$ output of the DBS, we place a thin lens (focal $f_1$) that images the output of the photon source in the plane of detector $D_1$ with a magnification $M=s_2/(s_0+s_1)$. The detector $D_b$ is placed on the optical axis. Using formula (\ref{GHF3}), one can calculate the three-photon amplitude in the detector planes. If a photon is detected by $D_b$ at time $t_*$, the wave function of the remaining $\lambda_1$ photons is projected on
\begin{equation*}
\begin{split}
&a(\mathbf{r}_a,t_a,\mathbf{r}'_a,t'_a)=  \delta(t_a-(t_*+\tau)) \ \delta(t_a-t'_a) \frac{\exp\left(-\frac{x_a^2+y_a^2}{4S^2M^2}\right)}{(2\pi S^2)^{1/2}} \\ & \ \delta(x_a-x'_a)  \ \delta(y_a-y'_a) \ \exp\left({\i 2\pi \frac{x_a^2+y_a^2}{\lambda_1 s_2}(1+1/M)}\right) \\ & h_2(0,-\mathbf{r}_a/M).
\end{split}
\end{equation*} 
where $\tau=(s_1+s_2-s_3-s_4)/c$ and $h_2(\mathbf{r}_b,\bm{\rho}_3)$ is the propagator from the nonlinear device to the detector $D_b$. In the plane of detector $D_a$, the photons exhibit spatial bunching. The field is a linear superposition of localized two-photon Fock states of light. Due to the magnification factor $M$, the spatial extension of that linear superposition can be much larger than $S$. Such a quantum state of light is interesting in the context of coherent super-resolution imaging \cite{Giovannetti2009}: The object to be imaged would be placed in the plane of detector $D_a$ in order to be illuminated with that special quantum state. Resolution enhancement only matters in optical systems in which geometrical aberration have been eliminated. Wavefront curvature must also be under control to avoid distortions due to non-isoplanatism \cite{Brainis2009d}. The scheme of Fig. \ref{fig:2} makes it possible to control the wavefront curvature of the $\lambda_1$ photons by tailoring the detection of the $\lambda_2$ photon. For instance, one can make the wavefront of $\lambda_1$ photons flat in the plane of $D_a$ by placing a lens (focal $f_2$) in the path to $D_2$ (see Fig. \ref{fig:2}) and choosing $s_3$, $s_4$ and $f_2$ such that $2(\lambda_2/\lambda_1)M(M+1)=s_2/((1/f_2-1/s_4)^{-1}-(s_0+s_3))$. This solution exists if the inequalities $s_4(s_0+s_3)/(s_0+s_3+s_4)<f_2<s_4$ are satisfied.

In summary, we developed a formalism that allows us to analyse many interesting issues related to the propagation of arbitrary number states of light using the wave function formalism. We derived a generalization of the Huygens-Fresnel principle that accounts for the propagation of field correlations (including entanglement) in space and time and showed how to applied it in practice. The formalism is very helpful to design sources of heralded $N$-photons states with engineered spatial profile.
 
This research was supported by the Interuniversity
Attraction Poles program, Belgium Science Policy,
under grant P6-10 and the Fonds de la Recherche Scientifique - FNRS (F.R.S.-FNRS, Belgium).

\end{document}